%% file: main.tex
\documentclass[conference]{IEEEtran}
\IEEEoverridecommandlockouts
\usepackage{cite}
\usepackage{amsmath,amssymb,amsfonts}
\usepackage{graphicx}
\usepackage{textcomp}
\usepackage{xcolor}
\def\BibTeX{{\rm B\kern-.05em{\sc i\kern-.025em b}\kern-.08em
    T\kern-.1667em\lower.7ex\hbox{E}\kern-.125emX}}

\usepackage[T1]{fontenc}
\usepackage[utf8]{inputenc}
\usepackage{listings}
\usepackage{algorithm}
\usepackage{algorithmicx}
\usepackage{algcompatible}
\usepackage{algpseudocode}
\usepackage{adjustbox}
\usepackage{multirow}
\usepackage{subfig}
\usepackage{placeins}
\usepackage{hyperref}
\usepackage{url}
\usepackage{color}

\algrenewcommand\alglinenumber[1]{\scriptsize #1}

\definecolor{gray97}{gray}{.97}
\definecolor{gray75}{gray}{.75}
\definecolor{gray45}{gray}{.45}
\newcommand{\blindtext}{0}
    
\begin{document}
\title{Dynamic reconfiguration for malleable applications using RMA
\if \blindtext 0
\thanks{This research was funded by the project PID2023-146569NB-C22 supported by MICIU/AEI/10.13039/501100011033 and ERDF/UE. Researcher I.~Martín-Álvarez was supported by the predoctoral fellowship ACIF/2021/260 from Valencian Region Government and European Social Funds.}
\fi
}

\if \blindtext 0
\author{\IEEEauthorblockN{1\textsuperscript{st} Iker Mart\'in-\'Alvarez}
\IEEEauthorblockA{\textit{Dept. Ing. y Ciencia de Computadores} \\
\textit{Universitat Jaume I}\\
Castell\'on, Spain \\
ORCID: 0000-0002-3337-3298}
\and
\IEEEauthorblockN{2\textsuperscript{nd} Jos\'e I. Aliaga}
\IEEEauthorblockA{\textit{Dept. Ing. y Ciencia de Computadores} \\
\textit{Universitat Jaume I}\\
Castell\'on, Spain \\
ORCID: 0000-0001-8469-764X}
\and
\IEEEauthorblockN{3\textsuperscript{rd} Maribel Castillo}
\IEEEauthorblockA{\textit{Dept. Ing. y Ciencia de Computadores} \\
\textit{Universitat Jaume I}\\
Castell\'on, Spain \\
ORCID: 0000-0002-2826-3086}
}
\fi
\if \blindtext 1
\author{\IEEEauthorblockN{Blind Authors}}
\fi

\maketitle

\begin{abstract}
This paper explores the use of novel one-sided communication methods based on remote memory access (RMA) operations in MPI for the dynamic resizing of malleable applications, with the aim of minimizing the impact of data redistribution on application execution. 
After their integration into MaM, a library which provides features for converting applications into dynamic ones, these methods are compared with traditional collective-based approaches. 
In addition, we expand MaM with a strategy that allows background redistribution to be performed using one-sided communications. 
The results show that the proposed method is unable to surpass the collective communication due to the way in which the new method is initialized. This issue should be addressed in future work.
\end{abstract}





\input{sections/01-Intro}
\vspace{-5pt}
\input{sections/02-StateOfTheArt}
\vspace{-5pt}
\input{sections/03-Background}
\vspace{-5pt}
\input{sections/04-Method}
\vspace{-5pt}
\input{sections/05-Results}
\vspace{-5pt}
\input{sections/06-Conclusions}

\bibliographystyle{IEEEtran}
\bibliography{sn-bibliography}

\end{document}

%% file: sections/01-Intro.tex
\section{Introduction}
The exascale era of High Performance Computing (HPC) has seen a steady increase in system capabilities, driven by hardware advances, such as improved memory, storage, interconnects and processor parallelism, as well as new programming models, runtimes, and libraries. 
However, studies show that the efficient utilization of resources such as CPUs, GPUs, and memory remains challenging~\cite{Jie2023}. 
Thus, it is common for jobs to  underutilized resources during the execution, resulting in wasting energy.
These inefficiencies often leave nodes idle while other jobs wait, highlighting the need for the system to be managed more efficiently.

Dynamic resource management (DRM) allows jobs to adjust the amount of resources allocated to them during runtime, as long as the Resource Manager System (RMS) and the applications can handle these changes. 
This capability has proven to be an effective strategy to optimize the use of HPC infrastructures according to different objectives. 
Usually, the main objectives are to maximize resource utilization~\cite{Chadha2021}, increase computational~\cite{sergiothesis} or energy efficiency~\cite{Iserte2019a,Cascajo2023}, and improve I/O performance~\cite{Sanchez2023}.

From an application perspective, this adaptability is known as malleability~\cite{Feitelson1996}, defined in this paper as the ability of a distributed parallel application to dynamically adjust its size by changing the number of MPI~\cite{mpi40} ranks allocated to it at any point during its execution.
This flexibility improves application performance by scaling up when resources are available and scaling down when demand is high, thereby enabling optimal resource utilization.
Malleability is activated at specific checkpoints within the application, and requires the execution of a series of stages.
\begin{enumerate}
    \item \textit{Reconfiguration feasibility}: The RMS decides whether to resize the job according to a dynamic resource allocation policy. 
    If not, the subsequent steps are not performed.
    
    \item \textit{Process management}: The RMS assigns a number of cores to the job, determining how many MPI ranks are created or terminated. 
    Processes that exist before resizing are considered \textit{sources}, while those that continue after resizing are considered \textit{drains}.
    
    \item \textit{Data redistribution}:  At this stage, data are transferred from \textit{sources} to \textit{drains}, which usually dominates the reconfiguration cost.
    
    \item \textit{Resume execution}. The application resumes execution with the \textit{drains}.
\end{enumerate}



This paper focuses on the data redistribution stage in MPI applications, introducing two methods based on the MPI Remote Memory Access (RMA) one-sided model. 
Our objective is to determine whether this approach outperforms the two-sided schemes previously studied ~\cite{exampi23} by lowering redistribution costs and thereby reducing DRM overheads.
%

Unlike traditional two-sided communication, where the sender and receiver actively coordinate each exchange, one-sided communication can reduce costs by avoiding explicit synchronization.
Additionally, \textit{source} processes are freed up, as they do not need to participate actively~\cite{Dinan2016}, allowing them to continue executing the application while communication proceeds.
To ensure correctness without interrupting redistribution progress, we extend a prior synchronization strategy in which \textit{drains} confirm transfer completion.
 
We compare the RMA one-sided model with the baseline method, a two-sided alternative defined in~\cite{exampi23}, using a synthetic benchmark through Proteo~\cite{proteo_2024}, to measure redistribution time.

Following these comments, the main contributions of this paper are the following:
\begin{itemize}
    \item 
    Design of two data redistribution methods based on the RMA one-sided model.
    \item 
     Design of the synchronization strategy \textit{Wait Drains}, in which the \textit{sources} continue their execution until they confirm that the \textit{drains} have finished receiving the data.
    \item Evaluation of the developed techniques, comparing their performance with those presented in previous studies.
\end{itemize}

The following of the paper is as follows. Section~\ref{sec:stateart} introduces similar approaches for data redistribution in DRM. Section~\ref{sec:background} showcases the tool Proteo, used for the evaluation of the proposed methods. Section~\ref{sec:method} explains the proposed RMA methods for data redistribution. Section~\ref{sec:results} presents an evaluation of the methods against a previous one. Lastly, Section~\ref{sec:conclusion} summarizes the paper and shows the future work.

%% file: sections/02-StateOfTheArt.tex
\section{Background}
\label{sec:stateart}
Checkpoint/restart (C/R) techniques were initially employed to support MPI process malleability.
These approaches store the state of a job in non-volatile storage so it can be reloaded when needed. 
In this view, malleability can be regarded as a C/R variant in which execution is halted and later resumed with a different process configuration. 
Therefore, individual processes are responsible for saving and restoring the appropriate data according to the drain process count~\cite{ElMaghraoui2007,Gupta,alya,Lemarinier2016}.

Traditional C/R solutions suffer from poor performance due to the high cost of disk access during read and write operations. More recent approaches adopt in-memory data redistribution, which offers higher performance at the expense of support for node-failure recovery. While this limitation is critical for C/R, it is less relevant in the context of malleability. Therefore, malleability-oriented solutions predominantly follow the in-memory approach.




Dynamic frameworks that support in-memory data redistribution have developed different strategies to address this stage. In~\cite{sergiothesis}, data redistributions are categorized in three ways:
\textit{Automatic}, \textit{Semiautomatic}, and \textit{Manual}. 
The following describes these categories together with some tools that utilize these techniques with different implementations:
\begin{itemize}
    \item \textit{Automatic}, 
    where the user data communication patterns among processes is not explicitly codified.
    For example, MaM~\cite{mam_api_2024} and Flex-MPI~\cite{Martin2015} provide generalist automatic data redistribution using data structure registers, whereas DMR~API~\cite{Iserte2018} leverages data dependencies.
    \item \textit{Semiautomatic} methods provide data redistribution for specific scenarios. 
    For instance, AMPI~\cite{ampi} is based on unique virtual memory addresses, whereas DMRlib~\cite{Iserte2020} provides predefined redistribution patterns. 
    Additionally, MaM also provides predefined redistribution patterns, as it may not be able to automatically redistribute certain complex data structures.
    \item In the \textit{manual} mode, users are fully responsible for defining the communication pattern  among processes. 
    Thus, coders must implement data redistribution directly with point to point or collective functions. 
    For example, Elastic~MPI~\cite{elasticmpi} does not provide any assistance with data redistribution.
\end{itemize}


%% file: sections/03-Background.tex
\section{Proteo}
\label{sec:background}
Proteo is a highly configurable framework for building benchmarks that assess how application malleability affects real-world workloads, enabling fair and direct comparisons across approaches~\cite{proteo_2024}. 
It consists of two modules: (i) the Synthetic Application Module (SAM), which emulates MPI applications from user-defined parameters; and (ii) the Malleability Module (MaM), which performs dynamic reconfigurations. 
Proteo also provides monitoring submodules that record performance metrics for both emulated and real applications.

MaM enables malleability providing combinable methods and strategies for coordinating process management and data redistribution. 
It also provides an API that streamlines the conversion of existing parallel applications into malleable~\cite{mam_api_2024}. 
In what follows, we focus on the MaM elements relevant to this work.

\textbf{Process management.} 
All methods~\cite{spawn_methods} operate by defining an initial group of $N_S$ \textit{sources} and a new group of $N_D$ \textit{drains}, allowing processes to temporarily belong to both groups during reconfigurations. 
In this study, the \textit{Merge} method is used, which spawns $(N_D - N_S)$ processes when $N_D > N_S$, or removes $(N_S - N_D)$ ones when $N_S > N_D$. 
This method was originally proposed in \textit{Flex-MPI}~\cite{Martin2015}, and it was adapted into MaM~\cite{spawn_par25} ensuring it suffers from no limitations from the existence of \textit{MPI\_COMM\_WORLD}.

\textbf{Data redistribution.} 
A key feature of MaM is  its ability to efficiently and automatically redistribute data between \textit{sources} and \textit{drains}.
It supports scalars and one-dimensional structures of primitive or MPI-derived types, with data classified as \textit{constant} (unchanged during execution and transferable via blocking or non-blocking operations) or \textit{variable} (changes during execution and requires blocking the application). 
This distinction is crucial in determining whether redistribution can be overlapped with computation using non-blocking primitives, or whether blocking operations must be used to preserve correctness. 
MaM implements several redistribution methods, which are described in~\cite{exampi23}. 
In this work, we use the \textit{Collective} method via \texttt{MPI\_Alltoallv} and enable three strategies: (i) \textit{Threading} (background redistribution via auxiliary threads), (ii) \textit{Non-blocking} (overlap transfers with computation), and (iii) \textit{Wait Drains} (a similar idea to \textit{ii}, but with an additional synchronization to ensure completion).

%% file: sections/04-Method.tex
\section{RMA Data redistribution}
\label{sec:method}
\subsection{RMA General Description}
RMA is a one-sided communication model introduced in MPI-2~\cite{Dinan2016}, that enables one rank (\textit{origin}) to directly read from or write to the memory of another rank (\textit{target}) without requiring immediate synchronization. 
This reduces communication overhead, making it particularly interesting for dynamic communication patterns. 
RMA is built around three key components: memory windows (shared memory regions), remote operations (\textit{Put}/\textit{Get}), and synchronization mechanisms that control access and ensure consistency through well-defined communication \textit{epochs}. 
An \textit{epoch} defines the time interval during which an \textit{origin} process is allowed to perform operations on the memory windows of \textit{target} processes.
Among the available memory access models, the passive model is especially suitable for redistributions, as it enables access to a target's memory without requiring its active participation. 
For a more in-depth explanation of how one-sided communication works, refer to the MPI standard~\cite{mpi40}.

Listing~\ref{code:def_mpi_rma} shows the primary functions for performing transfers using the passive model in the context of redistributions.
The \textit{Win\_create} and \textit{Win\_free} functions create and release a window, respectively. 
Both are collective, blocking operations for all processes within the associated communicator.
On the other hand, the \textit{Get} function allows data to be read from a window, whereas the  \textit{Lock} and \textit{Unlock} functions open and close an epoch on a specific \textit{target}.

\begin{lstlisting}[caption=MPI functions for passive one-sided communications., label=code:def_mpi_rma]
int MPI_Win_create(...);
int MPI_Win_free(...);
int MPI_Get(...);
int MPI_Win_lock(int locktype, int rank, int assert, MPI_Win win);
int MPI_Win_unlock(...);
int MPI_Win_lock_all(int assert, MPI_Win win);
int MPI_Win_unlock_all(...);
\end{lstlisting}
\vspace{-6pt}


\subsection{MaM's implementation}
\label{subsec:implementacionMaM}
Two additional methods have been incorporated into MaM to support data redistribution under the passive RMA model. 
In this model, sources (targets) expose memory windows containing the data to be redistributed, enabling drains (origins) to perform direct reads without active source participation. 
Although MPI allows multiple data structures to be exposed within a single memory window using displacement-based addressing, this approach greatly increases implementation complexity. For this reason, is created a dedicated window for each data structure to simplify offset management.
The following three algorithms outline the procedures used by these methods.  
Algorithm~\ref{code:redist_calculate} describes how  each \textit{drain} determines which data to read from which \textit{sources} for block-based redistribution. 
First, the algorithm obtains the total number of \textit{sources} (L1) and the data range assigned to the \textit{drain}  $\left[ini,end\right]$ (L2). 
For each \textit{source} $i$, the algorithm obtains that \textit{source}'s data range $[s\_ini,s\_end]$ (L7) and computes its intersection with the \textit{drain} range (L8).
If the intersection is empty, no data are read; otherwise, the number of elements is stored in \textit{counts}$\left[i\right]$ (L13–L15). 
The \textit{displs} array stores  the offsets in the \textit{drain} buffer where the elements will be written (L16).
\textit{first\_source} and \textit{last\_source} variables define the \textit{source} range with a non-empty intersection (L10, L19), while \textit{first\_index} gives the starting offset within \textit{first\_source} (L11); note that, under the block-based scheme, this offset is only required for the first window accessed. 
Finally, this procedure also allocates the per–data-structure memory requirements for each \textit{drain}, and the resulting allocations are returned to the user for deallocation.

\begin{algorithm}[tbh!]
\small
\begin{algorithmic}[1]
\State $s\_size$ = $Get\_source\_group ()$ 
\State $ini, end$ = $Block\_id\_d (myId)$
\State $counts$ = $calloc(s\_size)$
\State $displs$ = $calloc(s\_size+1)$
\State $first\_source$ = $-1$
\For {( $i=0$; $i < s\_size$; \textit{i++} ) }
    \State $s\_ini, s\_end$ = $Block\_id\_s (i)$
    \If {($ini < s\_end  \text{ }\&\&\text{ }  end > s\_ini$)}
        \If {($first\_source == -1$)}
            \State $first\_source = $\textit{i}
            \State $first\_index = ini - s\_ini$
        \EndIf
        \State $big\_ini$ = $ini > s\_ini ? ini : s\_ini$
        \State $small\_end$ = $end < s\_end ? end : s\_end$
        \State $counts[i]$ = $small\_end - big\_ini$
        \State $displs[i+1]$ = $displs[i] + counts[i]$
    \Else
        \If {($first\_source != -1$)}
            \State $last\_source = i$
            \State \textit{break}
        \EndIf
    \EndIf
\EndFor
\end{algorithmic}
\caption{\small Communication parameters on the \textit{drain} side.}
\label{code:redist_calculate}
\end{algorithm}



Algorithms~\ref{code:rma1} and~\ref{code:rma2} describe how \textit{drains} access the \textit{source} windows to read data. 
They are largely similar, but differ in that reads are either executed with one epoch by each access to a target process (Algorithm~\ref{code:rma1}) or within a single epoch (Algorithm~\ref{code:rma2}).
Additionally, both algorithms begin with \textit{sources} and \textit{drains} creating a window: the former expose it over their data buffer (L5 and L21 in Algorithm~\ref{code:rma1}), while the latter create it over an empty area (L3) in Algorithm~\ref{code:rma1}. 

In Algorithm~\ref{code:rma1}, the access to these windows is performed in shared mode by the drains (L9), under the assumption of no conflicts (L10), which allows multiple windows to be read in parallel. 
For each source to be accessed, the drain opens an epoch, posts the read request, and resets the indices so that the data are placed contiguously (L11-15).
Afterwards, it iterates  over the sources again to close the epochs and ensure local completion (L16-18). 
Ranks that are \textit{only sources} create and free their window (L21 and L23), without actively participating in the epochs. 
Finally, all processes free the created window (L19 and L23).


Conversely, in Algorithm~\ref{code:rma2}, a single epoch is created and opened by calling \textit{Win\_lock\_all} (L10).
Then, all \textit{Get} operations are posted (L11–L14), after which the process finally waits locally for completion (L15).

These approaches are mainly defined to assess whether the time taken to open and close epochs is significant.

\begin{algorithm}[tbh!]
\small
\begin{algorithmic}[1]
\If {(rank is \textit{drain})}
    \If {(rank is only \textit{drain})}
        \State $data = NULL$
    \EndIf
    \State $window$ = \texttt{MPI\_Win\_create}$(data)$
    \State $first\_source = get\_first\_source()$
    \State $last\_source = get\_last\_source()$
    \State $first\_index = get\_first\_index()$
    \State $lock = MPI\_LOCK\_SHARED$
    \State $assert = MPI\_MODE\_NOCHECK$
    \For{( $i=first\_source$; $i < last\_source$; \textit{i++} ) } 
        \State \texttt{MPI\_Win\_lock}$(i, lock, assert)$
        \State \texttt{MPI\_Get}$(i\to myId, displs[i], first\_index,counts[i])$
        \State $first\_index = 0$
    \EndFor
    \For{( $i=first\_source$; $i < last\_source$; \textit{i++}  ) } 
        \State \texttt{MPI\_Win\_unlock}$(i)$
    \EndFor
    \State \texttt{MPI\_Win\_free}$(window)$
\Else
    \State $window$ = \texttt{MPI\_Win\_create}$(data)$ 
    \State \ldots
    \State \texttt{MPI\_Win\_free}$(window)$
\EndIf
\end{algorithmic}
\caption{\small Method RMA1: Lock+Unlock.}
\label{code:rma1}
\end{algorithm}

\begin{algorithm}[tbh!]
\small
\begin{algorithmic}[1]
\If {(rank is \textit{drain})}
    \If {(rank is only \textit{drain})}
        \State $data = NULL$
    \EndIf
    \State $window$ = \texttt{MPI\_Win\_create}$(data)$ 
    \State $first\_source = get\_first\_source()$
    \State $last\_source = get\_last\_source()$
    \State $first\_index = get\_first\_index()$
    \State $assert = MPI\_MODE\_NOCHECK$
    \State \texttt{MPI\_Win\_lock\_all}$( assert )$
    \For{( $i=first\_source$; $i < last\_source$; \textit{i++} ) } 
        \State \texttt{MPI\_Get}$(i \to myId, displs[i], first\_index,counts[i])$
        \State $first\_index = 0$
    \EndFor
    \State \texttt{MPI\_Win\_unlock\_all}
    \State \texttt{MPI\_Win\_free}$(window)$
\Else
    \State $window$ = \texttt{MPI\_Win\_create}$(data)$ 
    \State \ldots
    \State \texttt{MPI\_Win\_free}$(window)$
\EndIf
\end{algorithmic}
\caption{\small Method RMA2: Lockall+Unlockall.}
\label{code:rma2}
\end{algorithm}


\subsection{Enhancements for background redistributions}
\label{subsec:WT}

Although the RMA model does not require active participation from the \textit{sources} during reads, the two previously proposed methods (Algorithms~\ref{code:rma1} and~\ref{code:rma2}) still exhibit blocking points that limit the overlap of computation and communication.
The most significant issue is that the operations of creating and freeing windows are collective and block all ranks.
As such, \textit{sources} cannot continue execution while \textit{drains} are reading.
Moreover, when a rank acts as both a \textit{source} and a \textit{drain}, invoking \textit{Win\_unlock} blocks until its own RMA operations are complete, stalling local computation. 
Next, we introduce two strategies to solve these challenges.

\subsubsection{Threading} 
In the first alternative, each \textit{source} uses an auxiliary thread to perform data redistribution in the background using either Algorithm~\ref{code:rma1} or~\ref{code:rma2}. 
This allows the main thread to continue executing while periodically checking for completion.


\subsubsection{Wait Drains} 
This is a global completion strategy based on a non-blocking barrier. 
In this strategy, \textit{drains} confirm \textit{sources} when their reads are complete, meanwhile \textit{sources} continue computing, and windows are unlocked and freed only once the non-blocking barrier has finished, ensuring that all ranks have completed their tasks. 
Three main modifications to the previous Algorithms~\ref{code:rma1} and~\ref{code:rma2} are introduced to implement the latter strategy.


First, to avoid the blocking behaviour of \texttt{Win\_free}, a global completion detector is added: all ranks post an \texttt{MPI\_Ibarrier} and \textit{sources} poll it with \texttt{MPI\_Test}, continuing execution while communication is in progress. 

Second, to avoid local blocking at \textit{MPI\_Unlock}, \textit{MPI\_Get} is replaced with \textit{MPI\_Rget}, which returns an \textit{MPI\_Request}. Thus, ranks acting as 
\textit{sources \& drains} periodically call \textit{MPI\_Test} to monitor progress and continue execution while the operation is ongoing.


Finally,  Algorithms~\ref{code:rma1} and~\ref{code:rma2} have been split into two parts to account for the previous changes: \texttt{Init\_RMA} and \texttt{Complete\_RMA,} which starts and ends the data redistribution, respectively. 

\begin{figure}[tb]
    \centering
    \includegraphics[width=0.9\linewidth]{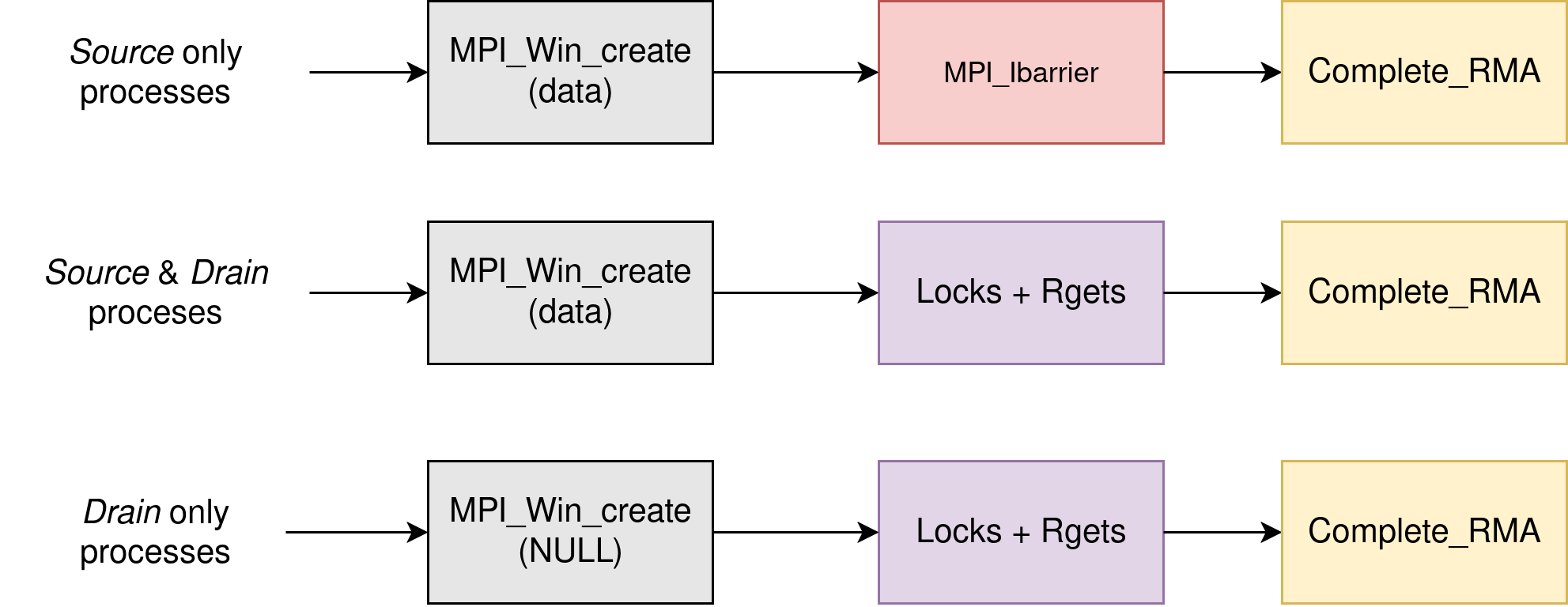}
    \caption{\texttt{Init\_RMA} flowchart for a background redistribution with RMA.}
    \label{fig:RMA-Init}
\end{figure}

Figure~\ref{fig:RMA-Init} shows the \texttt{Init\_RMA} flowchart, which initiates communication.
For more information on specific operations, refer to Algorithms~\ref{code:rma1} (L1–L15 and L21) and~\ref{code:rma2} (L1–L14 and L18).
As with the algorithms, the diagram distinguishes three execution paths according to process role: \textit{drain}-only, \textit{source}-only, and ranks that act as both. 
All sources create a window to expose data (NULL for \textit{drain}-only). 
\textit{Source}-only ranks do not perform reading operations, so they simply call \textit{MPI\_Ibarrier} to notify the others that their reads have finalized.
\textit{Drains} first acquire a lock on the target window with \textit{Win\_lock} and then post their reads with \textit{Rget}. 
Finally, all of them proceed to \textit{Complete\_RMA} to complete the operation.

\begin{figure*}[tb]
    \centering
    \includegraphics[width=0.7\linewidth]{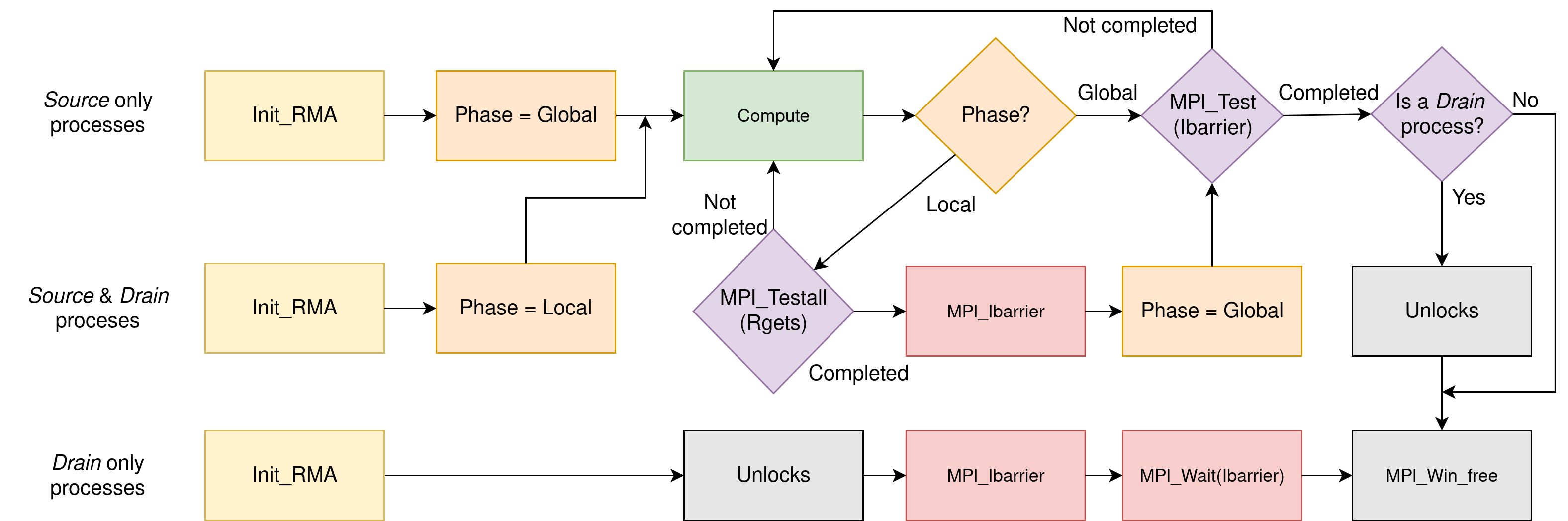}
    \caption{\texttt{Complete\_RMA} flowchart for a background redistribution with RMA.}
    \label{fig:RMA-Complete}
    \vspace{-10pt}
\end{figure*}

Figure~\ref{fig:RMA-Complete} presents the flowchart for the \textit{Complete\_RMA} function. 
This enables asynchronous data redistribution without blocking \textit{source} processes, while distinguishing between different execution paths according to process roles.
The flow of \textit{Complete\_RMA} is organized into two phases: a local phase, in which each rank acting as a \textit{drain} completes its pending reads, and a global phase, in which all ranks synchronize to close and free the windows.

\textit{Drain}-only ranks wait for their reads to complete during the local phase using (\texttt{Win\_Unlock}). 
They then enter the global phase, participate in the global barrier (\texttt{MPI\_Ibarrier}), and finally free the windows (\texttt{Win\_free}).
Note that these processes may block while waiting for the operation to complete.

\textit{Source}-only ranks have no local phase and therefore enter the global phase directly by posting \texttt{MPI\_Ibarrier}.
They continue computing while the barrier is incomplete, polling its status with \textit{MPI\_Test}.
Once the barrier completes, they free the window with \textit{MPI\_Win\_free}.

\textit{Source+drain} ranks first execute the local phase, overlapping computation and communication while waiting for their remote reads to complete (\texttt{MPI\_Testall}). 
Once this phase is done, they post \texttt{MPI\_Ibarrier} and enter the global phase, continuing computations while polling with \texttt{MPI\_Test} until all ranks have finished.
Finally, they free the windows using \texttt{MPI\_Win\_free}.




%% file: sections/05-Results.tex
\section{Experimental results}
\label{sec:results}

\subsection{\textit{Hardware} and \textit{Software}}
The experiments were conducted on an 8-node cluster, each node featuring two 10-core Intel Xeon 4210 CPUs (160 cores total) and connected via a 100 Gbps InfiniBand EDR network. 
MPICH 4.2.0\footnote{Dynamic process spawning is not supported with MPICH using the netmod UCX. 
Additionally, there are some limitations with the OFI netmod and Infiniband, but the configuration indicated works.} 
with CH4:OFI and the provider \textit{verbs} (InfiniBand) was used. 
\if \blindtext 0
Both the Proteo version\footnote{\url{https://lorca.act.uji.es/gitlab/martini/malleability_benchmark/-/tree/Sarteco25}} and experimental results~\cite{martin_alvarez_2025_15139406} are publicly available.
\fi
\if \blindtext 1
Both the Proteo version\footnote{\url{https://anonymous.4open.science/r/PDP26-Proteo-BB3C/README.md}} and experimental results~\cite{results_blind} are publicly available.
\fi

The main goal of this work is to evaluate the data redistribution stage by comparing the newly proposed methods with those previously incorporated into MaM’s.
To facilitate this analysis, all experiments used the same process management, the \textit{Merge} method in blocking mode. 

The experimental evaluation uses SAM to emulate the Conjugate Gradient algorithm~\cite{RBA94}.
The problem size is $72,067,110 \times 72,067,110$ with a total of $5,414,538,962$ non-zero elements, requiring approximately 64 GB of memory.
Each experiment performs a single reconfiguration from $NS$ to $ND$ processes, using 12 combinations from the set $\{20, 40, 80, 160\}$,
where a process pair $P=(NS \rightarrow ND)$ defines one such combination.
The number of nodes per run is set to $\lceil N/20 \rceil$ (with $N = \max(NS, ND)$) to optimize resource usage.
To ensure robust statistical analysis, each experiment was repeated 20 times, and the median result was taken as the representative value.

We compare three data redistribution methods: COL (collective), Algorithm~\ref{code:rma1} (RMA-Lock), and Algorithm~\ref{code:rma2} (RMA-Lockall), in both blocking and non-blocking modes. 
When run asynchronously, COL uses one of these strategies: \textit{Threading} (T), \textit{Non-Blocking} (NB), or \textit{Wait Drains} (WD). 
However, RMA-Lock and RMA-Lockall can only use T and WD.

NB and WD strategies are identical except for the communication completion criterion at the \textit{sources}. 
In NB, a source deems communication complete once it has sent all its messages, whereas in WD, completion is determined by notifications from the \textit{drains} (Section~\ref{subsec:WT}). 
Moreover, NB is not applicable to the RMA-based methods, since, in these methods, \textit{sources} only expose memory and cannot determine themselves when remote accesses have completed.

The set of non-blocking versions analyzed is the product $\mathcal{V}=M \times S$, where the method set is $M=\{\text{COL}, \text{RMA-Lock}, \text{RMA-Lockall}\}$ and the strategy set is $S=\{\text{NB}, \text{WD}, \text{T}\}$. 
Thus, each version is an ordered pair $V=(m,s)\in\mathcal{V}$. 
Section~\ref{subsec:tiempos_sinc} evaluates blocking versions. Section~\ref{subsec:non-blocking} focuses on the non-blocking strategies NB and WD, whereas Section~\ref{subsec:threading} analyzes the threading strategy $T$, which is included in $\mathcal{V}$ for completeness. 


\subsection{Blocking Data Redistribution}
\label{subsec:tiempos_sinc}

\begin{figure}[tb]
    \centering
    \includegraphics[width=0.95\linewidth]{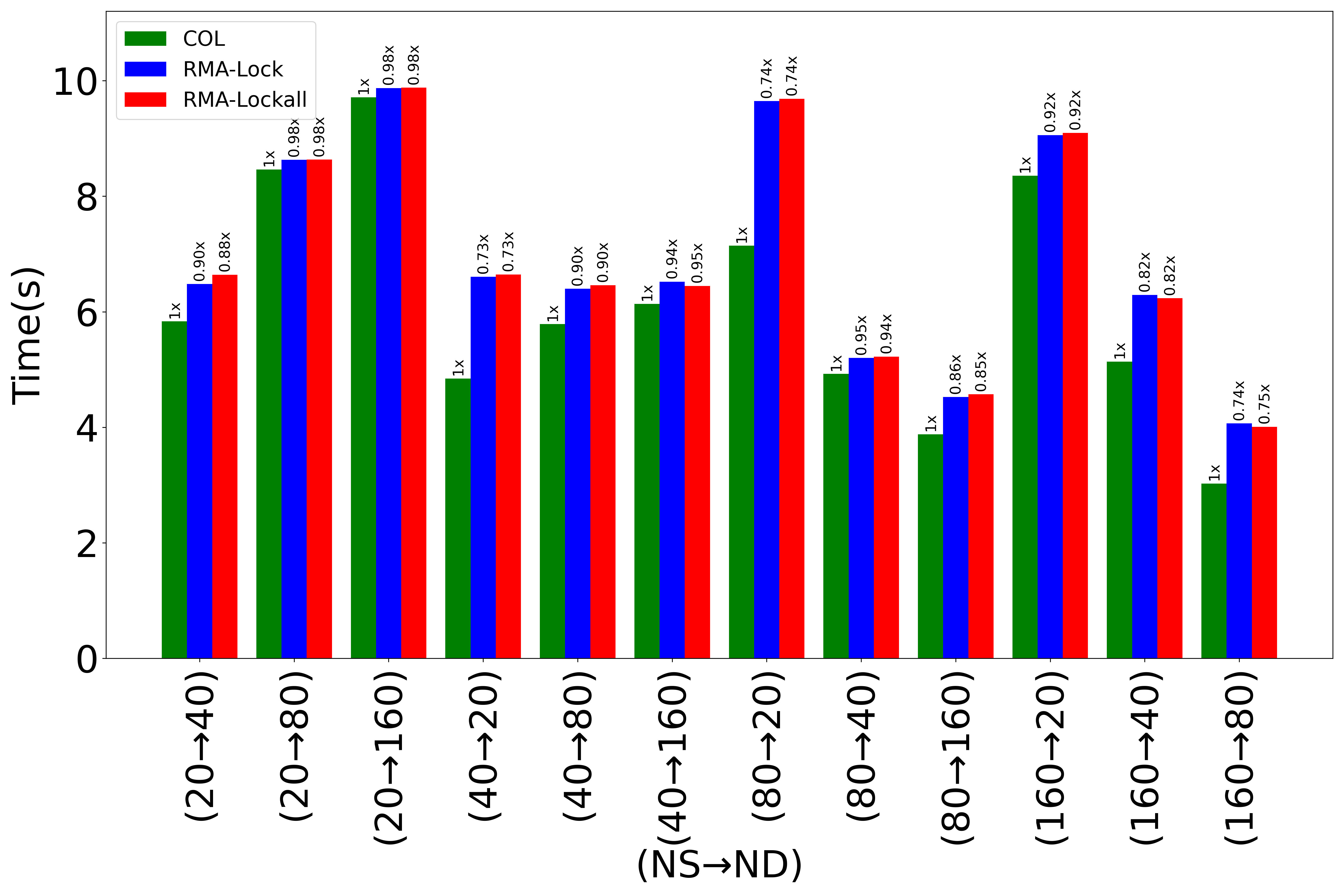}
    \vspace{-6pt}
    \caption{Reconfiguration times(s) in blocking versions, showing the speedups relative to COL for each pair ($NS \rightarrow ND$).}
    \label{fig:Time_S}
    \vspace{-10pt}
\end{figure}

Figure~\ref{fig:Time_S} shows the redistribution time (in seconds) for the blocking versions, as the number of \textit{sources} and \textit{drains} varies. 
For each configuration ($NS \rightarrow ND$), the  speedup relative to the first bar (COL) is also reported. 
RMA-Lock and RMA-Lockall exhibit nearly identical performance, with a maximum difference in speedup of only $0.02\times$. 
However, both consistently underperform compared to COL, with relative values ranging from $0.73\times$ to $0.99\times$. 
This overhead mainly arises from the collective cost of memory-window initialization. 
Overall, RMA-based methods are not recommended for blocking reconfigurations, as they are slower and more complex to implement than COL.

\subsection{Non-Blocking methods}
\label{subsec:non-blocking}

\if(0)
This subsection analyzes the behavior of non-blocking versions. 
We consider the method set $M=\{\text{COL}, \text{RMA-Lock}, \text{RMA-Lockall}\}$ and the strategy set $S=\{\text{NB}, \text{WD}, \text{T}\}$. 
The set of all versions is the product $\mathcal{V}=M \times S$; each version is an ordered pair $V=(m,s)\in\mathcal{V}$. 
This section focuses on the non-blocking strategies NB and WD; Section~\ref{subsec:threading} analyzes the threading strategy $T$, which is included in $\mathcal{V}$ for completeness. 
It is also required to define the process pair $P=(NS \rightarrow ND)$, where $NS$ and $ND$ are the numbers of \textit{source} and \textit{drain} processes in the reconfiguration.
\fi


Comparing different versions is not as simple as minimizing data redistribution time, since application behavior during the redistribution must also be considered. 
For this analysis, let $N_{\mathrm{it}}^{V,P}$ denote the number of iterations completed by version $V$ during the the data redistribution for pair $P$. We then store, for each $P$, the maximum iteration count achieved across all versions, as shown in Equation~\ref{eq:test1}:

\begin{equation}
  M^{P} = \max_{V \in \mathcal{V}} N_{it}^{V,P}
  \label{eq:test1}
\end{equation}
For simplicity, it is assumed that the application always performs more than $M^{P}$ iterations.

Equation~\ref{eq:test2} defines the data redistribution cost for a given version $V$ and pair $P$, as the sum of the redistribution time  $R^{V,P}$ and the time needed to reach the maximum iteration count $M^P$ from Equation~\ref{eq:test1}. 
If version $V$ completes the redistribution in fewer than $M^P$ iterations, the remaining iterations must be carried out at the per-iteration cost with  $ND$ processes, denoted as $T_{it}^{ND}$:

\begin{equation}
  f(V,P) = R^{V,P} + T_{it}^{ND}\,\bigl(M^P - N_{it}^{V,P}\bigr)
  \label{eq:test2}
\end{equation}
%
%
%
Finally, Equation~\ref{eq:test3} selects, for each pair $P$, the version $V$ that minimizes the data redistribution cost:
\begin{equation}
  V^*(P) = \arg\min_{V \in \mathcal{V}} f(V,P).
  \label{eq:test3}
\end{equation}
%
Using this approach, all versions are compared based on the time required to reach a given iteration, considering both the data redistribution time and the iterations needed to reach $M^P$. 
As such, the approach shows which version advances faster in the application.

\begin{figure}[tb]
    \centering
    \includegraphics[width=0.95\linewidth]{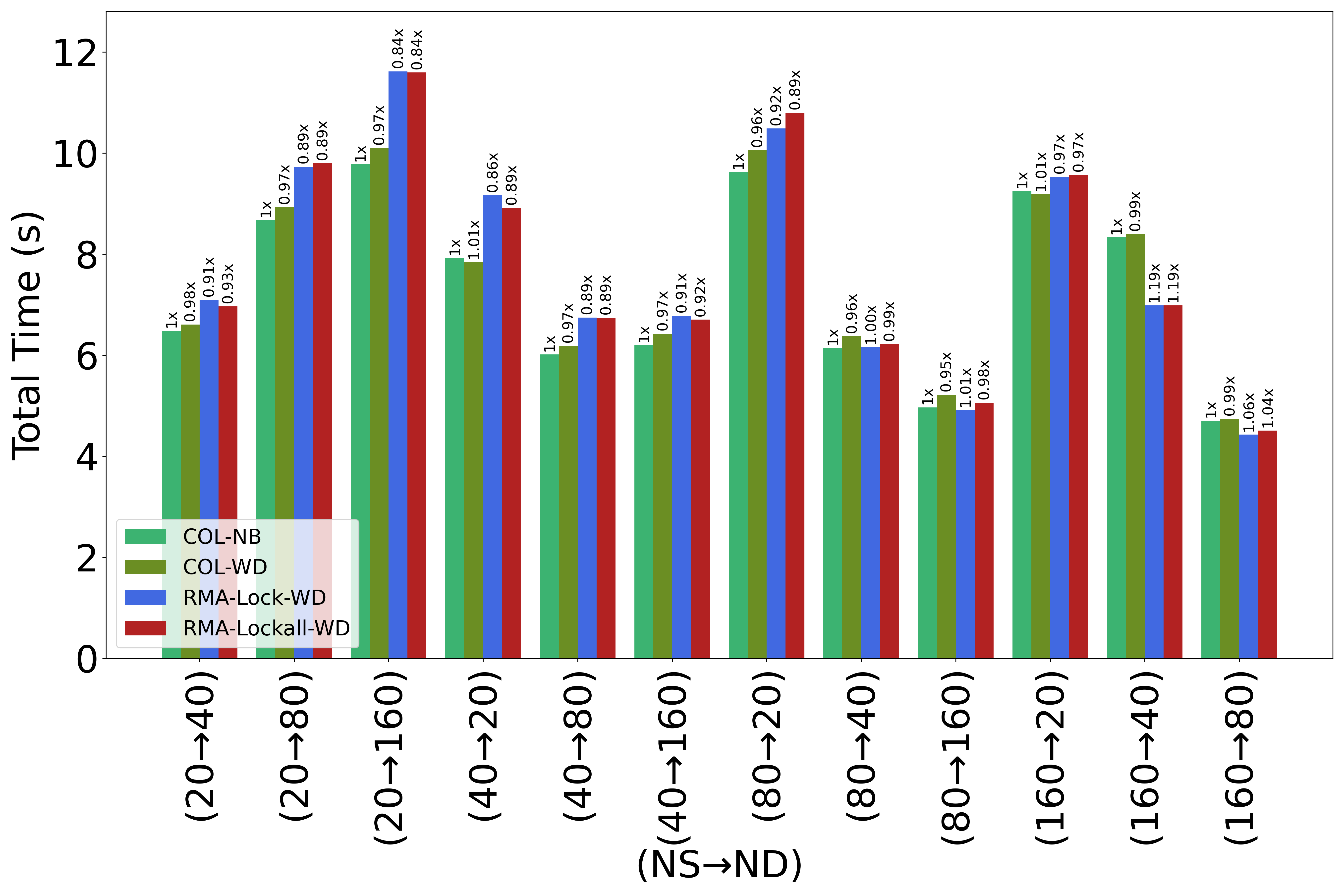}
    \vspace{-6pt}
    \caption{Total time after applying Equation~\ref{eq:test2} for NB and WD versions. The speedups relative to COL-NB are also shown for each pair ($NS \rightarrow ND$).}
    \label{fig:EqNB}
    \vspace{-10pt}
\end{figure}

Figure~\ref{fig:EqNB}  reports the corresponding times for each version, as well as the speedup relative to the first bar (baseline) for each pair $P$.
In all cases, COL-NB outperforms the other options and performs similarly to COL–WD. 
RMA variants provide a benefit only for $(160\rightarrow40)$ pair, achieving a $1.19\times$ speedup. 
Based on these results, COL is the preferred choice.
 

The following analysis examines the impact of concurrent data redistribution on the performance of source processes, providing additional insight into  the results presented in Figure~\ref{fig:EqNB}. 
The ratio $\omega = T_{\text{bg}}/T_{\text{base}}$, where $T_{\text{bg}}$ is the per-iteration time with background redistribution (overlapped execution) and $T_{\text{base}}$ is the baseline per-iteration time (without redistribution), quantifies the effect of overlapping data redistribution on the application's performance.

\begin{figure}[tb]
    \centering
    \includegraphics[width=0.95\linewidth]{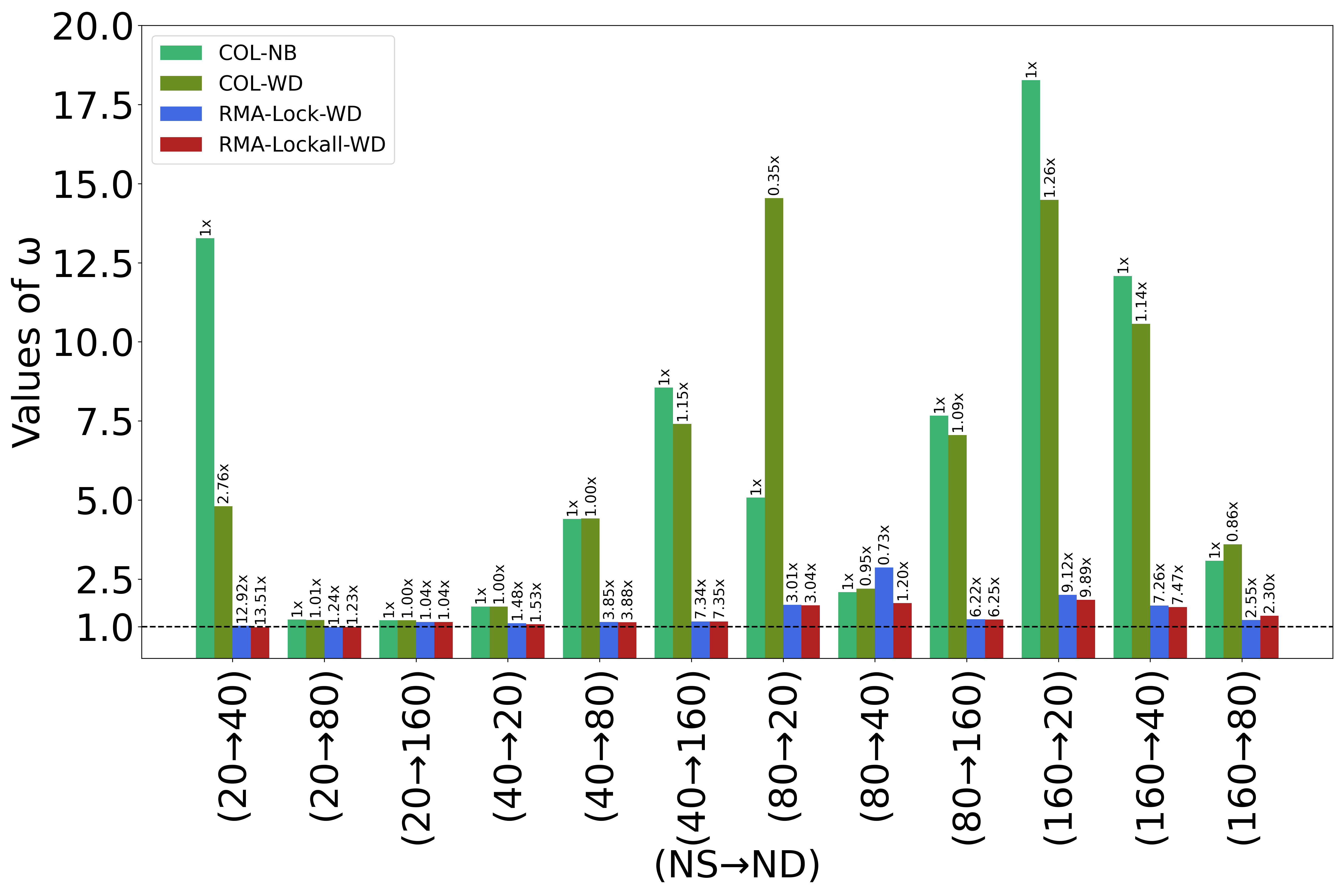}
    \vspace{-6pt}
    \caption{Relative increment of execution time when redistribution runs in background, represented by $\omega$, for NB and WD versions.}
    \label{fig:Omega_NB}
    \vspace{-10pt}
\end{figure}

Figure~\ref{fig:Omega_NB} shows the $\omega$ values for NB and WD. 
RMA-based methods produce the smallest slowdowns, with $\omega \approx 1$ in most cases and a worst case of $2.8$.  
This efficiency results from involving fewer processes in communication, since the \textit{source} processes only participate in creating and freeing  windows, and from avoiding the waiting time associated with global synchronization thanks to the use of \texttt{MPI\_Ibarrier}.  
Within these methods, RMA-Lockall–WD slightly outperforms RMA-Lock–WD by reducing the number of synchronization epochs. 
The largest $\omega$ values occur when the number of \textit{drains} is reduced ($160 \rightarrow 20$), likely due to increased contention in those scenarios.

\begin{figure}[tb]
    \centering
    \includegraphics[width=0.95\linewidth]{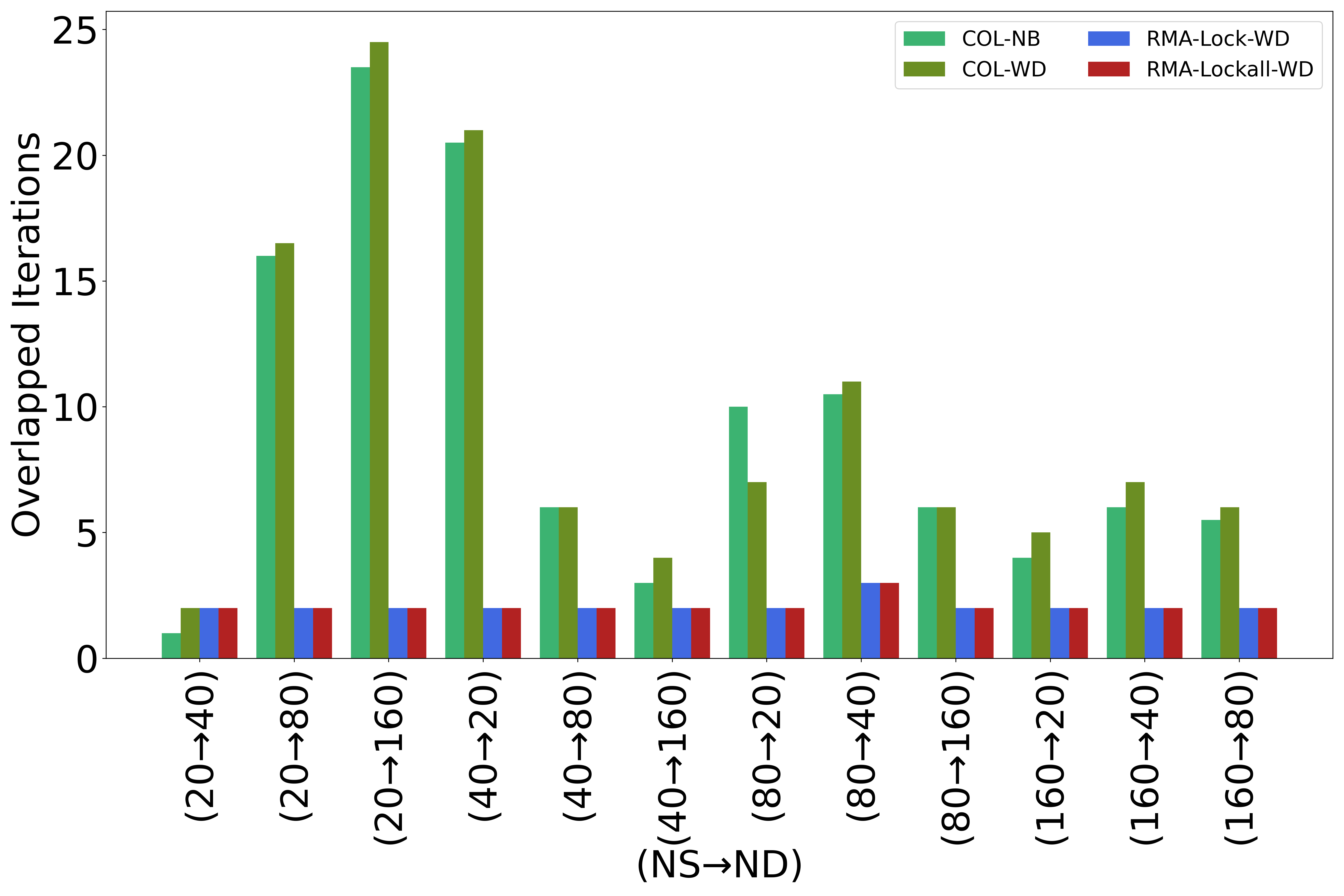}
    \vspace{-6pt}
    \caption{Total iterations when redistribution runs in background for NB and WD versions.}
    \label{fig:Iters_NB}
    \vspace{-10pt}
\end{figure}

Figure~\ref{fig:Iters_NB} reports the number of iterations that overlap with background redistribution across versions $V$ and process pairs $P$.
The analysis shows that COL variants exhibit more overlap, peaking at 24 for the ($20 \rightarrow 160$) pair due to the high communication overhead of the collective operation, in which all processes must actively participate. 
In contrast, RMA variants require only 2–3 iterations, since their dominant cost is the collective creation of memory windows across \textit{sources} and \textit{drains}. 
Some reads are also started during this creation, to the point that many of them are already completed by the time all windows are created. 
Therefore, fewer overlapping iterations are required.

Thus, the impact of background data redistribution on application time depends on two factors: the value of $\omega$ and the number of iterations executed while redistribution runs in background. 
Based on this, one might expect COL to be more expensive, since it exhibits larger $\omega$ and more overlapping iterations. 
However, Figure~\ref{fig:EqNB} shows the opposite: COL methods perform better. 
The decisive factor is the overhead of RMA, the collective creation of memory windows across sources and drains, which is substantially more expensive than the subsequent data reads. 
In fact, most read operations complete during the successive creation of the memory windows, resulting in a low effective number of overlapping iterations. 
Consequently, $\omega$ and the iteration count alone do not fully explain the observed costs; the window-creation overhead is the determining factor.


\subsection{Threading methods}
\label{subsec:threading}
This subsection examines auxiliary-thread support for the data redistribution stage. We implement COL, RMA-Lock, and RMA-Lockall as threaded variants (-T) and analyze them separately from the baseline versions.

\begin{figure}[tb]
    \centering
    \includegraphics[width=0.95\linewidth]{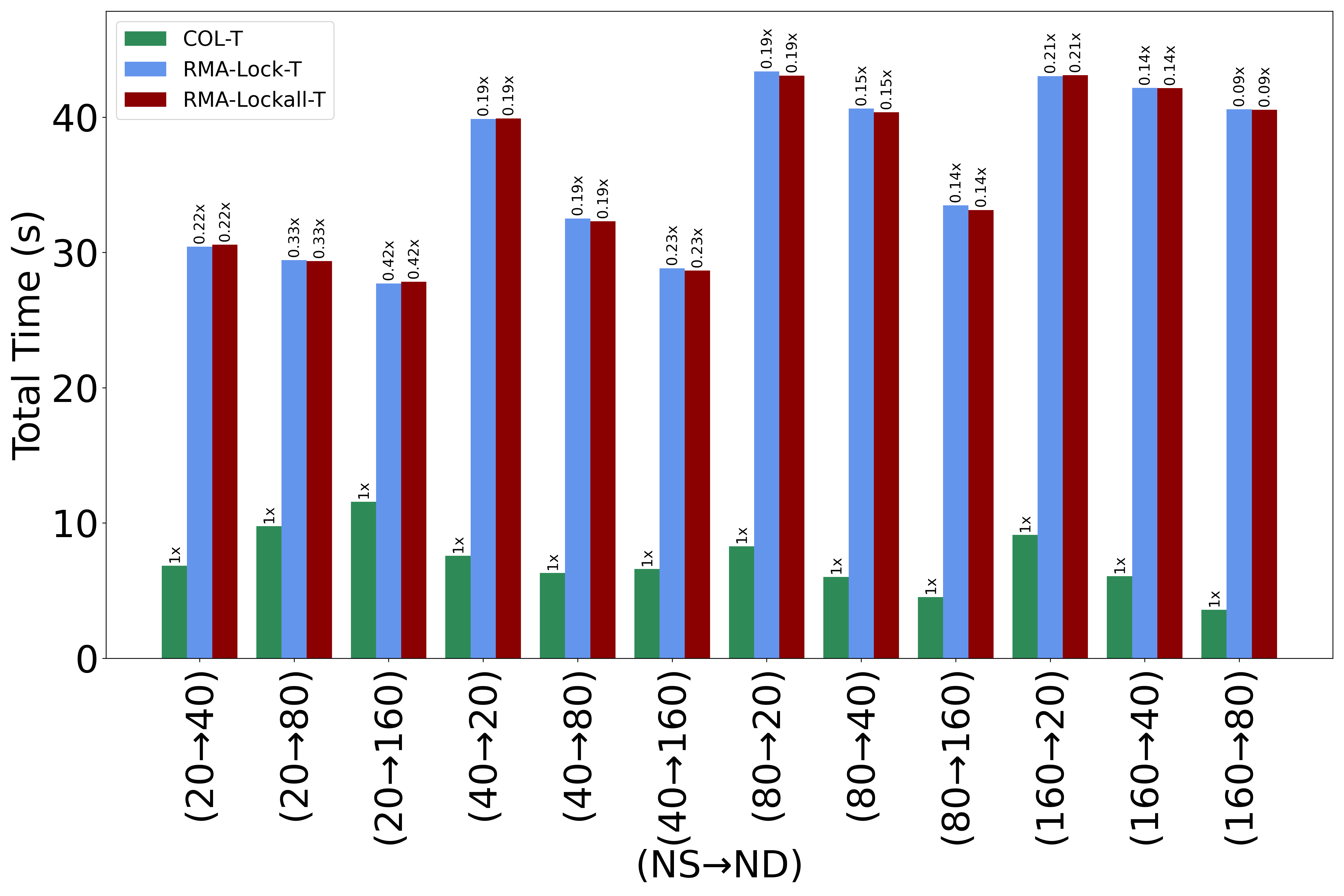}
    \vspace{-6pt}
    \caption{Total time after applying Equation~\ref{eq:test2} for T versions. The speedups relative to COLA are also shown for each pair ($NS \rightarrow ND$).}
    \label{fig:EqT}
    \vspace{-10pt}
\end{figure}

Figure~\ref{fig:EqT} applies Equation~\ref{eq:test2} to the T versions, showing the execution time and the speedups relative to the first bar for each pair $P$.
In all cases, COL-T outperforms the other variants, whereas RMA variants deliver substantially lower performance, with speedups ranging from $0.09\times$ to $0.42\times$. 
The effect of oversubscription introduced by the creation of an auxiliary thread is markedly greater for the RMA-based variants, as analyzed below.

\begin{figure}[tb]
    \centering
    \includegraphics[width=0.95\linewidth]{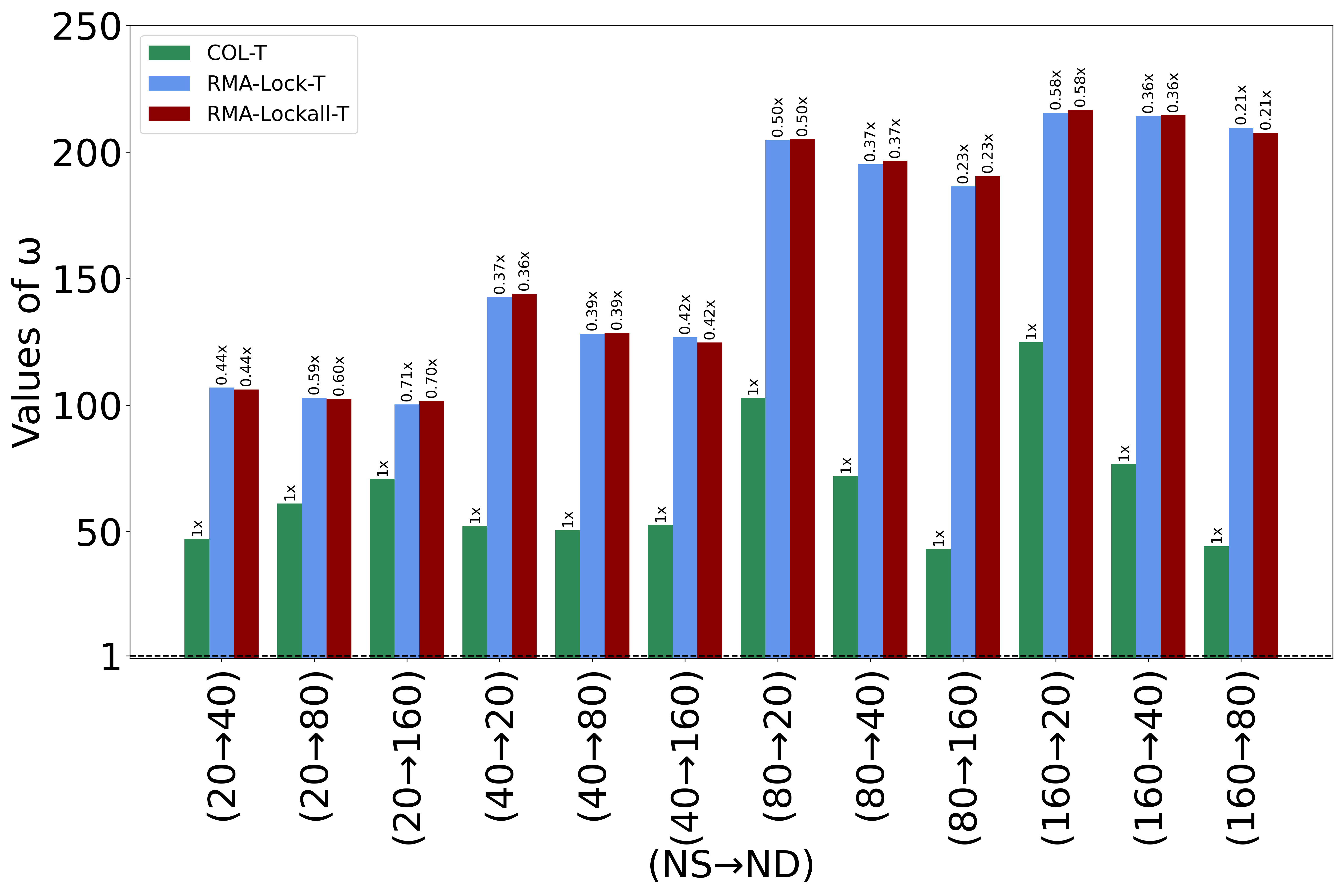}
    \vspace{-6pt}
    \caption{Relative increment of execution time when redistribution runs in background, represented by $\omega$, for T versions.}
    \label{fig:Omega_T}
    \vspace{-10pt}
\end{figure}

Figure~\ref{fig:Omega_T} reports the $\omega$ values.
They are high in all cases, exceeding 100 for the RMA methods and ranging from 43 to 123 for COL. 
A careful accounting of the iteration counts during redistribution provides the clearest explanation.


\begin{figure}[tb]
    \centering
    \includegraphics[width=0.95\linewidth]{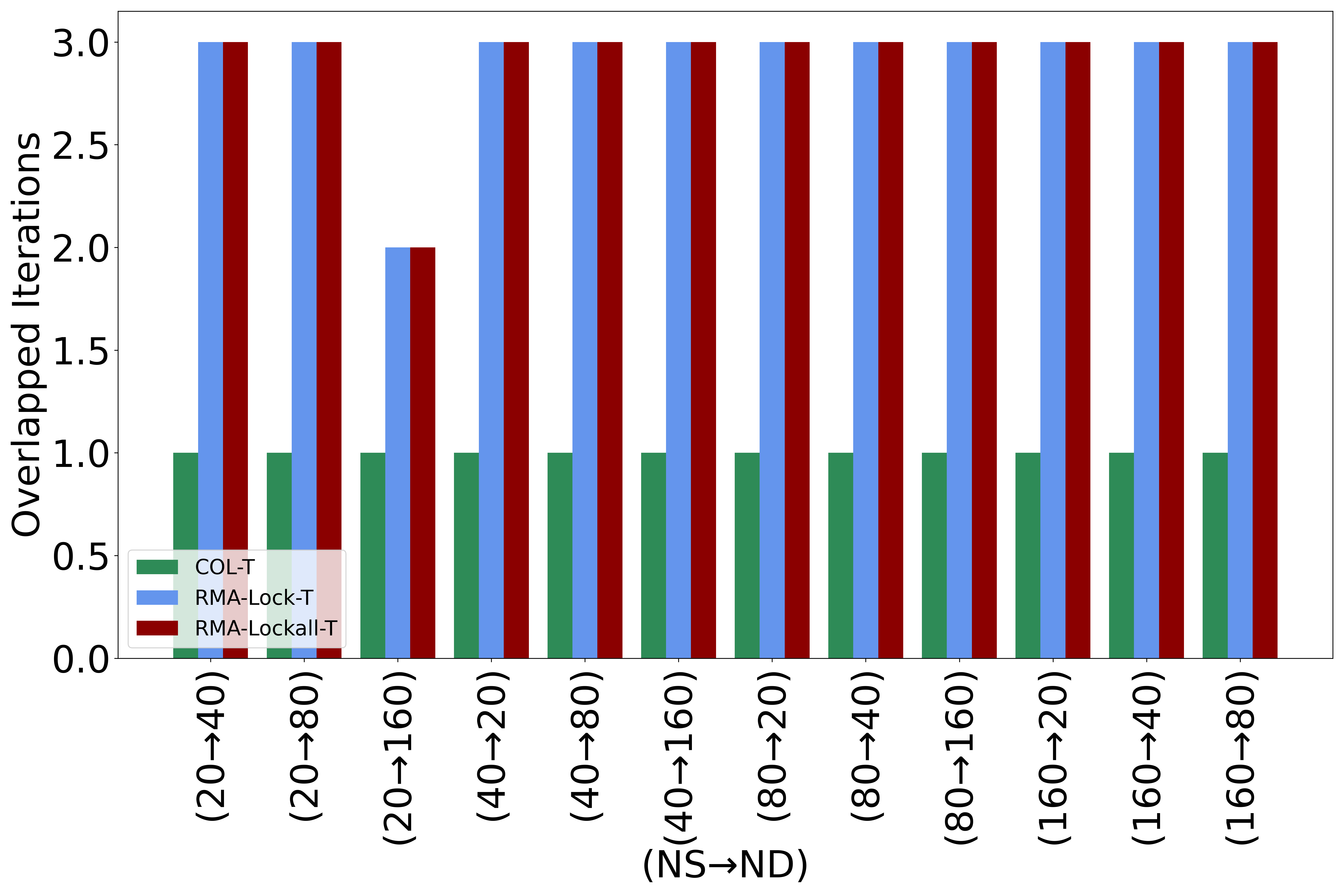}
    \vspace{-6pt}
    \caption{Total iterations when redistribution runs in background for T versions.}
    \label{fig:Iters_T}
    \vspace{-10pt}
\end{figure}

Figure~\ref{fig:Iters_T} reports the number of iterations that overlap with background redistribution.
The COL-T version completes only a single overlapping iteration across all pairs $P$. 
This occurs because the main thread of the sources running the application (CG) blocks in their first collective operation (\texttt{MPI\_Allgather}) until the auxiliary thread finishes its communications. 
Although multithreading (\texttt{MPI\_THREAD\_MULTIPLE}) has been enabled, the environment does not support it, which appears to reflect a limitation of the MPICH version used.
In contrast, the RMA variants typically require three iterations, but the cost per-iteration is $\ge 100\times$, making it too expensive. 
Therefore, even though the COL variant does not operate as intended, the RMA variants are not feasible due to the high per-iteration cost.

%% file: sections/06-Conclusions.tex
\section{Conclusions}
\label{sec:conclusion}

This paper extends MaM with six new data redistribution methods: two blocking and four executed as a background task. 
All of them are based on the one-sided communication model and employ the \textit{Threading} or \textit{Wait Drains} strategies. 
These enhancements enable background redistribution, allowing applications to continue running without active participation from \textit{sources}.


The primary motivation for the one-sided design was to allow the \textit{sources} to continue iterating without actively participating in the redistribution. 
However, the experiments show that these methods do not outperform the collective approach, even when using asynchronous techniques. 
The main reason is the high overhead associated with memory-window creation, a limitation that will be addressed in future work. 
On the positive side, the one-sided approach does not interfere with the application execution while a background redistribution is in progress. 
In summary, one-sided communication does not impact application performance during background resizing, but introduces a substantial initialization cost.

Future work will focus on reducing the cost of memory–window initialization. 
One approach is to create a single dynamic window in each \textit{source}, to which all memory that must be communicated is attached. 
Additionally, the communication volume for the \textit{Merge} method will be minimized by allowing \textit{sources} and \textit{drains} to retain as much data locally as possible.
Addressing these challenges will mitigate the dominant cost of data redistribution during reconfigurations and, consequently, reduce DRM costs.
